\documentstyle[prc,aps,psfig]{revtex}

\def\Journal#1#2#3#4{{#1} {\bf #2}, #3 (#4)}

\def\NP{{\em Nucl. Phys.}}
\def\NPA{{\em Nucl. Phys.} A}
\def\NPB{{\em Nucl. Phys.} B}

\def\PRL{\em Phys. Rev. Lett.}
\def\PRD{{\em Phys. Rev.} D}
\def\PRC{{\em Phys. Rev.} C}

\def\PR{{\em Phys. Rep.}}
\def\CPC{{\em Comp. Phys. Comm.}}

\def\JCP{{\em J. Comp. Phys.}}
\def\JEPT{{\em JEPT}}
\newcommand{\comment}[1]{}

\newcommand{\AmS}{{\protect\the\textfont2
  A\kern-.1667em\lower.5ex\hbox{M}\kern-.125emS}}

\newcommand{\etal}{{\it et al.}}
\begin{document}
\twocolumn[\hsize\textwidth\columnwidth\hsize
           \csname @twocolumnfalse\endcsname

\draft

\title{A Study of Parton Energy Loss in Au+Au Collisions at RHIC using Transport Theory }

\author{Y.\ Nara}
\address{ RIKEN-BNL Research Center, Brookhaven National Laboratory, Upton, NY }
\author{S.E.\ Vance}
\address{ Physics Department, Brookhaven National Laboratory, Upton, NY }

\author{P.\ Csizmadia}
\address{RMKI Research Institute for Particle and Nuclear Physics, 
	Budapest, Hungary}

\maketitle

\begin{abstract}
Parton energy loss in Au+Au collisions at RHIC energies is studied by
numerically solving the relativistic Boltzmann equation 
for the partons 
including $2 \leftrightarrow 2$ and 
$2 \rightarrow 2 + \mbox{final state radiation}$ collision processes.
%A numerical solution of the relativistic Boltzmann equation for a 
%distribution of partons with both 
%$2 \leftrightarrow 2$ and $2 \rightarrow 2 + \mbox{final state 
%radiation}$ parton processes is used to study energy loss in Au+Au 
%collisions at RHIC energies.  
Final particle spectra are obtained using 
two hadronization models; the Lund string fragmentation and 
independent fragmentation models.  
Recent, preliminary $\pi^0$ transverse momentum distributions 
from central Au+Au collisions 
at RHIC are reproduced using gluon-gluon scattering cross sections
 of $5-12$ mb, 
depending upon the hadronization model.  
Comparisons with the HIJING jet quenching algorithm are made.   
\end{abstract}
\vskip2pc]

%\narrowtext
%\begin{multicols}{2}

%\section{Introduction} 
Exciting, preliminary $\pi^0$ transverse momentum distributions from 
Au+Au collisions at $\sqrt{s_{NN}} = 130 \; \mbox{GeV}$ at RHIC have 
been recently reported~\cite{phenixPi02,phenixPi0}.    
When normalized to the mean number of binary collisions, 
the ratio of the central to the peripheral 
%$\pi^0$ transverse momentum 
distributions  
%divided by the binary collision normalized peripheral $\pi^0$ transverse 
%momentum distribution, 
%both normalized by the mean number of binary collisions,
%the resulting ratio 
increases up to $p_T \sim 2 \; \mbox{GeV/c}$, 
reaches a maximum value that is less than 1 and then decreases 
for $p_T > 2 \; \mbox{GeV}$.   
This preliminary data reveals a suppression of the production of $\pi^0$s
in central Au+Au collisions for $p_T > 2 \; \mbox{GeV/c}$.

The preliminary $\pi^0$ transverse momentum distribution has 
been reproduced by several 
calculations that included jet quenching~\cite{Wang:2001gv,LevaiQM}.   
Jet quenching occurs 
when a high energy jet passes through a medium and looses energy 
from the induced non-abelian radiation.  
Recently, intense theoretical activity has been devoted to calculating 
the energy loss of a fast parton traveling through 
QCD media~\cite{bdmps,zakharov,wiedemann,gyulassy}.
It has been shown that transverse momentum distributions of hadrons
at large $p_T$ are sensitive to the total energy loss of the fast 
partons~\cite{Wang:1992xy}. 
Jet quenching has been proposed as one of the signals of the formation of a 
quark-gluon plasma (QGP)~\cite{Wang:1992xy,Gyulassy:1994hr,Gyulassy:1990ye}.
Recent pQCD motivated calculations that include the nuclear effects 
of shadowing and energy loss via a modified fragmentation function 
have shown that the $\pi^0$ data can be 
reproduced with a constant energy loss of 
$dE/dz = 0.25 \; \mbox{GeV/fm}$~\cite{Wang:2001gv}. 
Using a similar approach, but with a $dE/dz$ that depends upon 
the number of collisions, it was shown that the data can also be
reproduced with $\bar{n} = 3-4$ average number of scatterings~\cite{LevaiQM}. 

While there have been many theoretical studies of jet quenching, few 
simulation models have incorporated this physics.
The HIJING simulation model~\cite{hijing} and the two other models which 
use the HIJING algorithms, such as HIJING/B\=B~\cite{Vance} 
and AMPT~\cite{ampt2} incorporate a simple jet quenching mechanism.
The HIJING jet quenching algorithm~\cite{hijing} 
assumes a simple gluon splitting scheme
with a fixed energy loss $dE/dz$.  The energy loss for gluon jets is 
twice that of the quark jets.  A jet
can only interact with locally comoving matter (strings) in the transverse 
direction and the points of the interactions are determined by the probability
$dP(l) = dl/\lambda e^{-l/\lambda},$
where $\lambda$ is the mean free path and $l$ is the distance the jet has
traveled between collisions.   When an interaction occurs (with 
a string), the medium induced radiation is simulated by forming a 
gluon kink in the string with $\Delta E = l dE/dz$ of the jets energy.
A jet can interact with the surrounding medium until 
it exits the system or its $p_T$ fall below a certain $p_0$ cut-off.
As the energy of a nuclear collision increases, more partons are 
produced, and this jet quenching algorithm leads to the production 
of many additional low energy gluons (gluon kinks). 
In a default HIJING jet quenching calculation with $dE/dz = 2$ GeV/fm, 
the charged hadron multiplicity per participant was shown to increase 
with the colliding energy
much faster than the data~\cite{Wang:2001bf,phobos1,phobos2}.
 
The effects of the parton scattering phase where only
elastic parton interactions are included
%using transport theory with only elastic parton scattering have 
have been studied with several models; ZPC~\cite{zpc}, 
AMPT~\cite{ampt2,ampt1}, MPC~\cite{mpc} and GROMIT~\cite{cheng}.
Calculations revealed\cite{ampt1} 
that elastic parton interactions only slightly 
decrease the $E_T$ and multiplicity and produce very little elliptic 
flow when typical gluon-gluon pQCD cross section of 
$\sigma_{gg} \sim 3$ mb and $dN_g/dy|_{y=0} \sim 200$ are used.
A large elliptic flow is only obtained when gross cross sections 
or large gluon densities are used~\cite{mpc}.  
Inelastic processes, like $gg \leftrightarrow ggg$, have been shown to be
be important~\cite{vni,chem,SMHWong} for thermalizing the partons
and in the radiative energy loss of the high energy jets. 

In this work, a model is introduced that describes 
the parton scattering phase for high energy heavy ion collisions
by numerically solving the Boltzmann equation for a distribution of partons 
with $2 \leftrightarrow 2$ and 
$2 \rightarrow 2 + \mbox{final state radiation}$ interactions.
The $2 \rightarrow 2 + \mbox{final state radiation}$ 
processes are needed to effectively simulate jet quenching 
and the dynamical scatterings of the 
partons lead to a reduction of the $E_T$ at mid-rapidity.
As a result, the incident energy dependence of the 
multiplicity per participant in this approach is similar 
to HIJING without jet quenching and is consistent with 
data~\cite{phobos1,phobos2}. 

%%%%%%%%%%%%%%%%%%%%%%%%%%%%%%%%%%%%%%%%%%%%%%%%%%%%%%%%%%%%%%%%%%%
%\subsection{model}
%%%%%%%%%%%%%%%%%%%%%%%%%%%%%%%%%%%%%%%%%%%%%%%%%%%%%%%%%%%%%%%%%%%

The initial parton distributions in our model are obtained 
from the HIJING event generator which
produces around 190 gluons at mid-rapidity for a typical 
Au+Au collision at $\sqrt{s_{NN}} = 130 \; \mbox{GeV}$. 
The nuclear shadowing effects in HIJING are included in all calculations
in this paper.   The system of 
partons is then evolved in time using the relativistic Boltzmann equation, 
\begin{equation}
  p^{\mu} \partial_{\mu} f(x,p) = C,
\end{equation}
where $f(x,p)$ is the distribution function
of the partons and $C$ is the collision integral.
For the collision integral, $2 \leftrightarrow 2$ and 
$2 \rightarrow 2 + \mbox{final state radiation}$ processes are included.
The distribution function, $f(x,p)$ is assumed to be the 
sum of the particles, 
\begin{equation}
  f(x,p) = {1\over N_{test}}\sum_{i=1}^{A N_{test}}
                         \delta(x_i - x) \delta(p_i - p),
\end{equation}
where $N_{test}$ is the number of ``test particles'' 
and $A$ is the actual number of partons produced in a collision.
The momentum of the partons is determined by HIJING and 
the space-time coordinates are calculated using simple uncertainty 
relations. 
% and the momentum of the partons as determined by HIJING. 
The formation time for partons is taken to be a Lorentzian
distribution with a half width $t = E/m_{T}^2$, where $E$ and $m_T$
are the parton energy and transverse mass, respectively~\cite{ampt1}.

%Number of test particles...
% rewrite this part... 
Monte-Carlo simulation is used to solve the Boltzmann equation.   
In this simulation, the on-shell partons are evolved in time along straight 
lines between collisions, where all of the collisions are time-ordered 
in a global frame.   After each collision, the flavor, position 
and momentum of the outgoing partons 
are determined and the list of possible collisions is updated.
% where all possible collisions are determined 
%and are time-ordered in a global frame.  
%Once a collision occurs, 
%the outgoing partons are determined and their collisions with other partons 
%are found.  
The space-time evolution of the system continues until there are no 
possible collisions between the partons. 

The ``parallel-ensemble'' method has been widely used to simulate
the low energy nucleus-nucleus collisions~\cite{Bertsch}.
In this method, collisions are determined using
the ``closest distance approach'', where  
a collision occurs if the minimum relative distance 
$b_{rel}$ for any pair of particles in their center of mass frame 
becomes less than interaction range as given by $\sqrt{\sigma/\pi}$.
Here, $\sigma$ is the total parton-parton cross section.
%for the pair in the center of mass.
 However, this method violates Lorentz invariance
due to the nature of action at a distance; a problem that has 
been studied by several authors~\cite{zpc,mpc,cheng,kwk}.
Two solutions to this problem have been proposed. 
The ``full-ensemble'' method replaces each particle 
by $N_{test}$ particles that interact with a reduced 
cross section $\sigma/N_{test}$~\cite{mpc,welke,lang}. 
In the limit of $N_{test} \to \infty$, one obtains the locality
in configuration space.   Another solution is the 
``local-ensemble'' method~\cite{localensemble,danielewicz}
where the probability for one pair of the test particles to
collide during the time interval of $\Delta t$ in the small volume
element $\Delta V$ is given by
\begin{equation}
  W = \sigma v \Delta t /(N_{test}\Delta V).
\end{equation}
Here $v$ is the relative velocity of the scattering particles.
In the limit of $\Delta V \to 0$, $\Delta t \to 0$, $N_{test} \to \infty$,
the solutions will converge to the exact solutions of the Boltzmann
equation.  In this study, the full-ensemble method is used with 
$N_{test}=6$.

% processes...
The parton-parton interactions included in this simulation are  
\begin{eqnarray}
 q + q' & \leftrightarrow & q + q', \quad
 g + g \leftrightarrow g + g, \quad  \\ \nonumber
 g + g & \leftrightarrow & q + \bar{q}, \quad
 g + q  \leftrightarrow  g + q. \quad
\end{eqnarray}
The cross sections for these processes are given by  
leading order (LO) perturbative QCD (LOpQCD)~\cite{pythia1} and 
explicitly take into account the running coupling constant 
$\alpha_s(Q^2)$. Here, the $Q^2$ scale is chosen to be the
squared transverse momentum transfer of the scattering process
$Q^2=p_T^2$ and
$\alpha_s(Q^2)$ is evaluated at the scale of 1 GeV$^2$, when $p_T < 1$ GeV.
Since these cross sections diverge when $p_T \to 0$,
a Debye screening mass $m_D=0.6$ GeV~\cite{zpc}
and quark thermal mass $m_q = 0.2$ GeV are introduced to regulate 
the propagators~\cite{kwk,SMHWong}. 
While these values should change with time, they 
are taken to be constant for simplicity. 
In this paper, two sets of results are obtained by multiplying these 
cross sections with two different factors ($K$-factors);
$K = 1.0$ and $K = 2.5$.
%instead of changing the Debye mass to see the cross section
%dependence for the parton evolutions.
For example, a $K$-factor of 1.0 yields
$\sigma_{gg} \sim 5.3\; \mbox{mb}$ for the gluon-gluon cross section,
$\sigma_{gq} \sim 2.0\; \mbox{mb}$ for the quark-gluon,
and $\sigma_{qq} \sim 0.5$ mb for the quark-quark,
and a $K$-factor of 2.5 yields 2.5 times the values for $K=1$.

% 2 -> 2 + FSR, including final state radiation
The $2 \rightarrow 2 + \mbox{final state radiation}$ 
processes are modeled 
using PYTHIA algorithms~\cite{pythia2}, where the two 
out-going partons are evolved with time-like branching taking 
into account angular ordering.  
%
%In a high energy parton-parton interactions, 
%partons can become off-shell and can radiate.
%To simulate effectively parton multiplication processes,
%the time-like branching is included using PYTHIA~\cite{pythia2} algorithms.
% include angular ordering %
%
%This approach reproduces gluon jet production in $e^+e^-$ interactions. 
This approach has been used to study
the gluon jet production in $e^+e^-$ interactions. 
During the branching, the life times of newly branched partons are obtained
using the uncertainly principle, $t_{form} \sim 1/Q$, where 
$Q$ represents the off-shellness of the parton in its
rest frame~\cite{eskwang}.
The minimum virtuality for the final state radiation is chosen 
to be 0.5 GeV~\cite{eskwang}.   The number of emitted gluons 
(amount of final state radiation) is sensitive to this cut-off parameter.  
Using a $Q^2 = 1 \; \mbox{GeV}^2$ with $K = 2.5$ yielded less final state 
radiation and hence less energy loss.   
A systematic study will be given later.
The maximum virtuality
in parton-parton scattering is assumed to be $Q_{max}^2=4p_{T}^2$,
where $p_T$ is the transverse momentum transfer of the scattering process.
Off-shell partons are not allowed to interact with other partons 
in the system.   While the final state (time-like) radiation is occurring, 
possible interactions with other partons are ignored,
in contrast to the parton
cascade model VNI~\cite{vni}.

%%%%%%%%%%%%%%%%%%%%%%%%%%%%%%%%%%%%%%%%%%%%%%%%%%%%%%%%%%%%%%%%%%%%%%%%%%
%  dn/dy gluons
%%%%%%%%%%%%%%%%%%%%%%%%%%%%%%%%%%%%%%%%%%%%%%%%%%%%%%%%%%%%%%%%%%%%%%%%%%
%
%\onecolumn
\begin{figure} %[b!]%Fig. 1.   
%\includegraphics[width = 9.0cm,keepaspectratio]{dndpt_glu.eps}
%\centerline{\psfig{figure=fig/dndy_parton.eps,width=6.5in,height=4in}}
\centerline{\psfig{figure=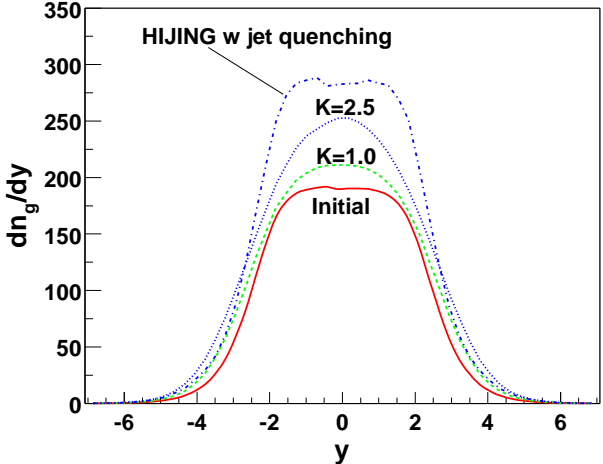,width=3.5in}}
\caption{ Rapidity distributions of mini-jet gluons
for $Au+Au$ collisions ($b<4.48$ fm) at $\sqrt{s_{NN}}=130$ GeV.
The solid line denotes the distribution from the initial condition
(HIJING no jet quenching),  and
the dashed and dotted line represent the distributions after rescattering
with a $K$ factor of 1.0 and 2.5 respectively.
The dotted-dashed line is the result from HIJING default calculation
($dE/dz=2.0$ GeV/fm).
}
\label{fig:dndyG}
\end{figure}

In Fig.~\ref{fig:dndyG},
the rapidity distributions of the mini-jet gluons for central Au+Au
collisions at $\sqrt{s_{NN}}=130 \; \mbox{GeV}$ are shown.
The solid curve represents the initial rapidity distribution of 
gluons obtained from the HIJING model without jet quenching;
$dN/dy_{g,y=0} = 190$.  With this initial distribution, 
the dynamical evolution of the partons with $2\rightarrow n$ interactions 
enhances the gluon multiplicity, where $dN/dy_{g,y=0} = 210$ 
for $K=1.0$, and $dN/dy_{g,y=0} = 250$ for $K=2.5$.
In comparison, HIJING with jet quenching with $dE/dz=2.0$ GeV/fm yields 280
gluons near mid-rapidity.  As observed, these values are sensitive 
to the magnitude of the parton-parton cross sections.

%%%%%%%%%%%%%%%%%%%%%%%%%%%%%%%%%%%%%%%%%%%%%%%%%%%%%%%%%%%%%%%%%%%%%%%%%%
%  dEt/dy partons
%%%%%%%%%%%%%%%%%%%%%%%%%%%%%%%%%%%%%%%%%%%%%%%%%%%%%%%%%%%%%%%%%%%%%%%%%%
\begin{figure}[b!]%Fig. 2.   
%\includegraphics[width = 9.0cm,keepaspectratio]{dndpt_glu.eps}
%\centerline{\psfig{figure=fig/dndy_parton.eps,width=6.5in,height=4in}}
\centerline{\psfig{figure=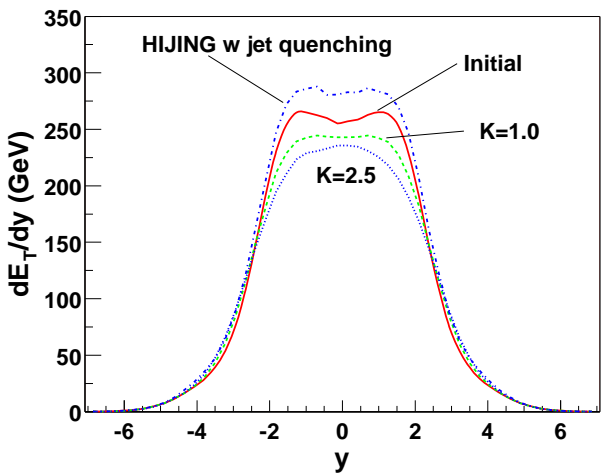,width=3.5in}}
\caption{ Transverse energy rapidity distributions of partons
for $Au+Au$ collisions ($b<4.48$ fm) at $\sqrt{s_{NN}}=130\; \mbox{GeV}$.
The solid line denotes the distribution from the initial condition and
the dashed($K=1.0$) and dotted ($K=2.5$) line represents
 the distributions after rescattering. HIJING with default jet quenching result
is shown by the dotted-dashed line.}
\label{fig:detdyParton}
\end{figure}
%\twocolumn
%%%%%%%%%%%%%%%%%%%%%%%%%%%%%%%%%%%%%%%%%%%%%%%%%%%%%%%%%%%%%%%%%%%%%%%%%%
% rapidity correlation
%%%%%%%%%%%%%%%%%%%%%%%%%%%%%%%%%%%%%%%%%%%%%%%%%%%%%%%%%%%%%%%%%%%%%%%%%%
\begin{figure}
\centerline{\psfig{figure=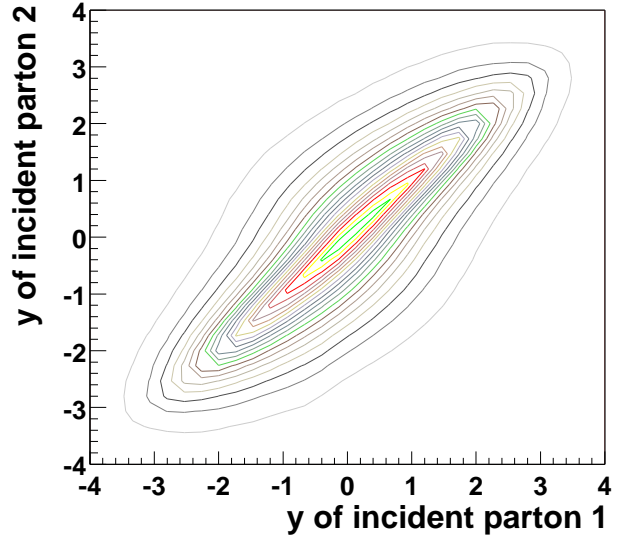,width=3.5in}}
\caption{ Rapidity correlation for parton-parton scattering for Au+Au
collisions ($b<4.48\; \mbox{fm}$) at $\sqrt{s_{NN}}=130\; \mbox{GeV}$.
$K=2.5$ is used.
}
\label{fig:yy}
\end{figure}

In Fig.~\ref{fig:detdyParton}, the rapidity distributions of mini-jet
transverse energy $dE_T/dy$ in central Au+Au collisions
at $\sqrt{s_{NN}}=130\; \mbox{GeV}$ is shown.
After the partons are dynamically evolved, the 
$dE_T/dy$ of the partons is reduced by approximately 13 GeV for $K=1.0$ 
and 27 GeV for $K=2.5$.  The loss in $E_T$  
results from the strong rapidity correlation in the parton-parton collisions.
The rapidity correlation for the case of $K=2.5$ is 
plotted in Fig.~\ref{fig:yy}.  
As the partons with similar rapidities collide,
the momentum is redistributed from transverse to longitudinal
and the $E_T$ decreases.  
The degree of the correlation depends on the parton-parton cross sections,
where the correlation becomes stronger as the cross sections decrease.
Most collisions in the simulation occur around the 
$p_T \sim 1.0\; \mbox{GeV/c}$.  
%
%When soft collisions of $p_T < 2\; \mbox{GeV/c}$ are not included in the
%simulation, the rapidity correlation disappears and there 
%is no decrease in the $E_T$.
%in our calculations
%and collisions with large rapidity gaps occur.
%Therefore, within our model, the collision with high $p_T > 2$ GeV/c
%is strongly suppressed dynamically.
%
In comparison, in Fig.~\ref{fig:detdyParton}, 
the HIJING jet quenching scheme increases the $dE_T/dy$.
In the HIJING jet quenching scheme, the medium induced radiation 
is simulated by forming a {\em collinear} gluon (kink in the string). 
In this approach, little of the transverse momenta is redistributed. 
%with some of the jets energy and the momentum of the particles are not
%strongly redistributed into the longitudinal direction.  

\begin{figure}
\centerline{\psfig{figure=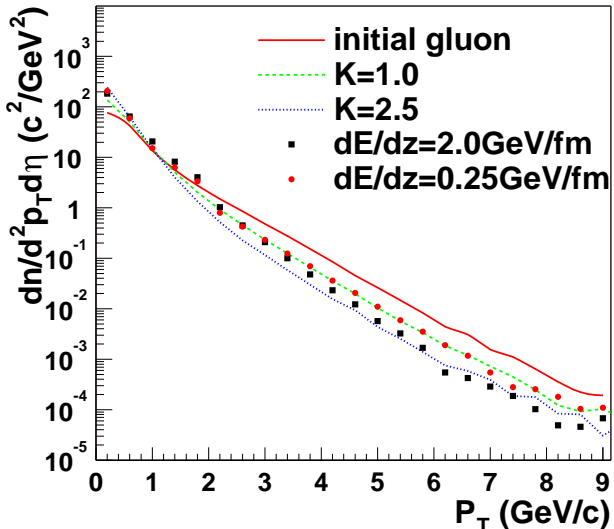,width=3.5in}}
\caption{ Transverse momentum distributions of gluons
for $Au+Au$ collisions ($b<4.48$ fm) at $\sqrt{s_{NN}}=130$ GeV.
The solid line denotes the distribution from the initial condition and
the dashed and dotted lines represent the distributions after rescattering
with $K=1.0$ and $K=2.5$, respectively.
HIJING results are also shown in squares ($dE/dz=2.0$ GeV/fm)
 and circles ($dE/dz=0.25$ GeV/fm).
}
\label{fig:dndptG}
\end{figure}

%%%%%%%%%%%%%%%%%%%%%%%%%%%%%%%%%%%%%%%%%%%%%%%%%%%%%%%%%%%%%%%%%%%%%%%%%%
% dn/dpt gluons
%%%%%%%%%%%%%%%%%%%%%%%%%%%%%%%%%%%%%%%%%%%%%%%%%%%%%%%%%%%%%%%%%%%%%%%%%%
Fig.~\ref{fig:dndptG} shows the gluon transverse momentum distributions
from our calculations at the $\sqrt{s_{NN}} = 130\; \mbox{GeV}$ for central 
Au+Au collisions together with HIJING jet quenching calculations.
The inclusion of $2 \rightarrow 2 + \mbox{final state radiation}$ processes 
% correct
%of the mini-jets
results in the reduction of high $p_T$ partons and 
a slight increase in the low $p_T < 1\; \mbox{GeV}$ region.
This effect is sensitive to the size of the parton-parton cross sections
and the assumed energy loss $dE/dz$. 

\medskip
%%%%%%%%%%%%%%%%%%%%%%%%%%%%%%%%%%%%%%%%%%%%%%%%%%%%%%%%%%%%%%%%%%%%%%%%%%
%  pi0 spectra lund
%%%%%%%%%%%%%%%%%%%%%%%%%%%%%%%%%%%%%%%%%%%%%%%%%%%%%%%%%%%%%%%%%%%%%%%%%%
Once the partons have finished interacting, 
two different hadronization models are used to obtain the 
hadron spectra; the Lund string fragmentation 
and independent fragmentation models.
% at the end of the time evolution of the partons.
The results of these two models are compared with one another and 
with the $\pi^0$ data.    

When using the Lund string fragmentation model~\cite{pythia2},
the string configurations are maintained throughout the evolution of the 
partonic system by determining the string configuration from 
the color amplitude of each parton-parton collision.  
In the PYTHIA manual~\cite{pythia1}, detailed explanations of the possible
string configurations for the relevant parton-parton scattering
processes are provided.
%
%   Although color connections are
%screened in a plasma, the string color connections are maintained for 
%the internal consistency of this calculation.  
%The string picture is adopted to model the soft component of the system
%and provides long-range knowledge of the position of the other jets.

\begin{figure}
\centerline{\psfig{figure=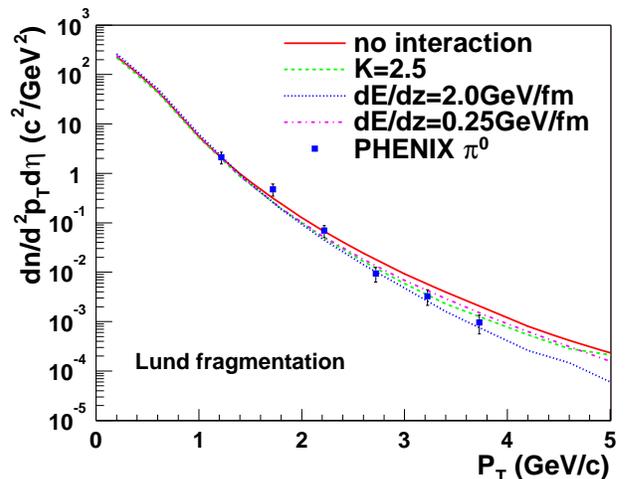,width=3.5in}}
\caption{ A comparison of neutral pion transverse momentum distributions
for central $Au+Au$ collisions ($b<4.48$ fm) at $\sqrt{s_{NN}}=130\; \mbox{GeV}$
with the PHENIX data~\protect\cite{phenixPi0} is presented.
The solid line denotes the distribution from the initial condition and
the dashed line represents the gluon distribution after rescattering
with the $K$ factor of 2.5.
The HIJING results for $dE/dz=2.0$ GeV/fm (dotted line)
and $dE/dz=0.25$ GeV/fm (dotted-dashed line) are also shown.
}
\label{fig:dndptPi0lund}
\end{figure}

As shown in Fig.~\ref{fig:dndptPi0lund},
the HIJING without jet quenching calculation (solid line)
overestimates the data at $p_T > 2 \; \mbox{GeV/c}$ as 
consistent with other pQCD motivated parton model 
calculations~\cite{phenixPi0,Wang:2001gv}.
Jet quenching is not seen in the low $p_T$ region, $1<p_T<2$ GeV/c,
as expected.
%
%This is a contrast to the pQCD parton model calculation, 
%which is overestimated
%in this region without jet quenching.
%The reason would be that in HIJING, soft interactions
%are modeled by string formation and it is well tested against the
%experimental data.
%However, pQCD parton model is applied down to $p_T \sim 1$ GeV/c.
%
As a result of the dynamical evolution of the partons,
the data can be reproduced with the $K=2.5$ cross section set. 
In addition, HIJING with jet quenching $dE/dz= 0.25 - 2.0$ GeV/fm is 
able to reproduce the data.   
The dominant contribution of the $\pi^0$
with $p_T \approx 2-4 \; \mbox{GeV/c}$ comes
from the hadronization of quarks with $p_T > 4 \; \mbox{GeV/c}$.  
Thus, the $\pi^0$ distribution is sensitive to the energy loss of the quarks
and it is very important to include the appropriate
$qq$ and $qg$ interactions.
    
%%%%%%%%%%%%%%%%%%%%%%%%%%%%%%%%%%%%%%%%%%%%%%%%%%%%%%%%%%%%%%%%%%%%%%%%%%
%  pi0 spectra indep. fragmentation
%%%%%%%%%%%%%%%%%%%%%%%%%%%%%%%%%%%%%%%%%%%%%%%%%%%%%%%%%%%%%%%%%%%%%%%%%%

\begin{figure}
\centerline{\psfig{figure=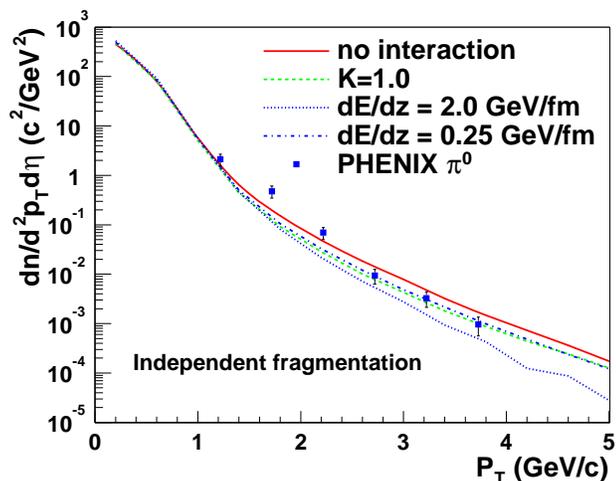,width=3.5in}}
\caption{ A comparison of neutral pion 
transverse momentum distributions 
for central $Au+Au$ collisions ($b<4.48$ fm) at $\sqrt{s_{NN}}=130\; \mbox{GeV}$
with the PHENIX data~\protect\cite{phenixPi0} is presented.
The solid line denotes the distribution from the initial condition and
the dashed line represents the distribution after rescattering
using $K=1$. 
The HIJING calculations (independent fragmentation model used)
 with jet quenching for $dE/dz=2.0$ GeV/fm (dotted line)
and $dE/dz=0.25$ GeV/fm (dotted-dashed line) are also shown.
%The $K=2.5$ gives us the same amount of underestimation
%of the data in HIJING result with $dE/dz=2.0$ GeV/fm.
}
\label{fig:dndptPi0indep}
\end{figure}

Since explicit local color connections are not believed to be maintained 
in a high density partonic system,
the independent fragmentation model is also used
to compute the final hadron spectra.   
The independent fragmentation model fragments all of the partons independently.
As compared to the Lund string model, this model does
not incorporate long range correlations between the partons. 
Within this hadronization scheme, the high $p_T$ $\pi^0$ data can be 
reproduced with $K=1.0$ as shown in Fig.~\ref{fig:dndptPi0indep}.
%correct
%We note here that 
Not shown is a calculation with $K=2.5$  
which had a larger energy loss, underestimating the data.
Since the long-range correlations between partons are absent,
the final hadron distribution strongly reflects the parton distribution.
Changing the fragmentation model in HIJING from the default Lund string 
model to the independent fragmentation model, 
the HIJING calculations with jet quenching show similar behavior.
Above $p_T > 2.0\; \mbox{GeV/c}$,
calculations assuming $dE/dz = 0.25\; \mbox{GeV/fm}$ are able to reproduce 
the data.
This value is the same as that obtained from the pQCD motivated parton model
calculations~\cite{Wang:2001gv}.

%%%%%%%%%%%%%%%%%%%%%%%%%%%%%%%%%%%%%%%%%%%%%%%%%%%%%%%%%%%%%%%%%%%%%%%%%%
%  discussion  about hadron yield
%%%%%%%%%%%%%%%%%%%%%%%%%%%%%%%%%%%%%%%%%%%%%%%%%%%%%%%%%%%%%%%%%%%%%%%%%%
%Recent measurements of the charged multiplicity per participant have
%been report by the PHOBOS collaboration~\cite{phobos2}.
%It was shown that HIJING {\it without} jet quenching is consistent with
%the energy dependence of charged hadron multiplicity per participant 
%near mid-rapidity, while HIJING with jet quenching 
%of $dE/dz = 2 \; \mbox{GeV/fm}$ increases much faster with 
%$\sqrt{s}$ than the data~\cite{phobos2}. 
%However, $\pi^0$ data at $\sqrt{s_{NN}}=130 \; \mbox{GeV}$
%can be account for with the rather wide range 
%of $dE/dz \approx 0.25 - 2.0$ GeV/fm in HIJING.
%One could explain the hadron multiplicity and $\pi^0$ data
%at the same time with different values of $dE/dz$
%The forthcoming data at $\sqrt{s_{NN}}=200$ GeV
%provides constraints on this effect.

The total hadron yields from our model near mid-rapidity 
are smaller than the HIJING {\em without} jet quenching values
by approximately 4\% when using Lund string fragmentation model 
and are larger by approximately 2\% when using the 
independent fragmentation model. 
We have checked that the model yields similar results at
$\sqrt{s_{NN}} = 200$ GeV. 
Our calculations are therefore consistent with data~\cite{phobos2} 
on the energy dependence of the hadron yield near mid-rapidity.

\medskip

%%%%%%%%%%%%%%%%%%%%%%%%%%%%%%%%%%%%%%%%%%%%%%%%%%%%%%%%%%%%%%%%%%%%%%%%%%
%  Summary and Conclusions
%\section{Conclusions}\label{sec::conclusion}
%%%%%%%%%%%%%%%%%%%%%%%%%%%%%%%%%%%%%%%%%%%%%%%%%%%%%%%%%%%%%%%%%%%%%%%%%%
In this paper, we study parton energy loss in central Au+Au collisions at 
RHIC by numerically solving the Boltzmann equation with 
$2\leftrightarrow 2$ and $2 \rightarrow 2 + \mbox{final state radiation}$ 
parton processes.
%Using the Lund string fragmentation model and the independent fragmentation
%model to obtain the final hadron spectra, 
Preliminary data on the $\pi^0$ $p_T$ distributions at RHIC were 
fit with gluon-gluon cross sections of 5-12 mb,
 depending upon the hadronization model. 
This approach provides the maximum incoherent bound for energy loss 
in comparison to the LPM effect.

With the given HIJING initial condition, the parton scatterings
reduce the parton $dE_T/dy|_{y=0}$ by 5-10\%. 
The magnitude of the change in the $E_T$ is
 sensitive to the scattering cross section. 
Interactions between particles that are strongly correlated in rapidity 
lead to a redistribution of the momenta from the transverse to the
longitudinal direction and are important in lowering the $E_T$
and the final hadron multiplicity. 
 Similar effects have been pointed out
in regard to the suppressed production of open charm
in the pre-equilibrium stage~\cite{openCharm}. 

To test the sensitivity of the $\pi^0$ spectra to the non-perturbative 
hadronization process, two different hadronization schemes were used.
Larger parton-parton scattering cross sections were needed to reproduce
the $\pi^0$ distributions when using Lund string fragmentation 
than when using independent fragmentation.  The spectra produced
by indepndent fragmentation are more sensitive to changes 
in the parton distributions, due to the lack of the long-range 
correlations introduced by the strings.  
These two different hadronization schemes therefore require different
amounts of parton energy loss in order to reproduce the data.

% The difference in the parton-parton
%cross sections resulting from the different hadronization methods 
%clouds the actual parton energy loss.  
% The difference in 
%correlations introduced by the strings.  The need for  
%larger parton-parton scattering cross sections when using the Lund 
%string fragmentation in order to reproduce the $\pi^0$ distribution 

%and clouds the actual energy loss of the partons.
%actual parton energy loss .     
%This difference introduces a factor of 2 in the results and clouds
%the actual parton energy loss. 
%As a result, the strength of the 
%energy loss depends partly upon the hadronization scheme.   

While we have investigated the energy loss in the parton scattering phase, 
future calculations should also address the change in the $p_T$ distributions 
of the particles from the late hadron gas stage.
The effect of the parton scattering phase using other initial 
conditions~\cite{pQCDsaturation,Kharzeev,MV,KV,ampt3} 
should also be explored.

% ?????????????????????????????????????????????????????????? % 
%We emphasize the importance of the rapidity correlations among
%interacting partons to suppress the hadron yield.
%The situation is rather similar with regard to the production
%mechanism of open charm discussed in Ref.~\cite{openCharm}.
%The space-momentum correlation in secondary parton scattering
%greatly suppresses the production of charm.
% ?????????????????????????????????????????????????????????? %

%In the present study, soft interactions are modeled by the string
%formation.  We do not consider the final state interactions among 
%those soft color field, because it has been suggested that
%the high partonic density produced from initial semi-hard scatterings
%screen the color field.
% However, string picture itself could be inadequate to describe
%the high density partonic system at RHIC~\cite{Kharzeev,MV,KV,ampt3}.
%Therefore it is interesting next study to explore the effect
%of the interactions among soft sector and hard sector
%replacing the soft sector to incoherent soft partons.

\medskip
We would like to thank S. Cheng, M. Gyulassy, P. Levai and I. Vitev for 
their helpful comments. 
This manuscript was authored under Contract No. DE-AC02-98CH10886 
with the U. S. Department of Energy and by Hungarian OTKA Grant T025579.

%%%%%%%%%%%%%%%%%%%%%%%%%%%%%%%%%%%%%%%%%%%%%%%%%%%%%%%%%%%%%%%%%%%%%%%%%%
% ref. 
%%%%%%%%%%%%%%%%%%%%%%%%%%%%%%%%%%%%%%%%%%%%%%%%%%%%%%%%%%%%%%%%%%%%%%%%%%

%\end{multicols}
\end{document}